# The Effects of Firing Conditions on the Properties of Electrophoretically Deposited Titanium Dioxide Films on Graphite Substrates

*Dorian Hanaor,[a,*] Marco Michelazzi[b], Jeremy Chenu[c], Cristina Leonelli[b], Charles Sorrell[a]*

a: University of New South Wales, School of Materials Science and Engineering, Kensington NSW 2052, Australia
b: University of Modena and Reggio Emilia, Department of Materials and Environmental Engineering, 41100 Modena, Italy
c: University of New South Wales, School of Chemcial Engineering , Kensington NSW 2052, Australia

*Corresponding author, email: dorian@ unsw.edu.au  Ph: 61-404-188810*

## Abstract

Thick anatase films were fabricated on graphite substrates using a method of anodic aqueous electrophoretic-deposition using oxalic acid as a dispersant. Thick films were subsequently fired in air and in nitrogen at a range of temperatures. The morphology and phase composition were assessed and the photocatalytic performance was examined by the inactivation of *Escherichia coli* in water. It was found that the transformation of anatase to rutile is enhanced by the presence of a graphite substrate through reduction effects. The use of a nitrogen atmosphere allows higher firing temperatures, results in less cracking of the films and yields superior bactericidal performance in comparison with firing in air. The beneficial effects of a nitrogen firing atmosphere on the photocatalytic performance of the material are likely to be a result of the diffusion of nitrogen and carbon into the $TiO_2$ lattice and the consequent creation of new valence band states.

Keywords: Photocatalysis; Films; $TiO_2$; Substrates; Sintering

## 1. Introduction

The development of novel approaches to water purification is of increasing importance as population growth and climate change place a growing strain on water resources [1, 2]. Photocatalysis is an attractive approach to water treatment as this technique does not involve the consumption of chemical reagents, enables the removal of a variety of pollutants, is effective across a wide range of pollutant concentration levels and can be achieved using solar irradiation as the sole energy input [3-5].

Owing to the distinct levels of its valence and conduction bands, $TiO_2$ has emerged as the leading material in photocatalytic applications [6, 7]. $TiO_2$ photocatalysis takes place through the photo-generation of an electron-hole pair, an exciton, by irradiation exceeding the band gap of the material. This leads to the generation of surface adsorbed radicals and subsequent oxidation of organic pollutants on $TiO_2$ surfaces [8]. The two phases of titanium dioxide most commonly used in photocatalysis are anatase and rutile. Despite the slightly larger band gap of anatase (~3.2eV vs. ~3.0eV), this phase is widely considered to exhibit superior photocatalytic activity as a result of greater levels of surface adsorbed radicals [9-12]. It has been frequently reported that mixed-phase $TiO_2$ exhibiting low levels of rutile alongside anatase exhibits enhanced performance through reduced electron-hole recombination [13-17].

As photocatalyzed destruction of pollutants takes place at close proximity to $TiO_2$ surfaces, a high surface area is advantageous for effective rates of pollutant removal [18]. For this reason studies of water purification by $TiO_2$ photocatalysts are often carried out using aqueous suspensions of powder [19, 20]. The disadvantage of using $TiO_2$ in the form of a powder suspended in the treated water is the required catalyst recovery processes, for this reason the immobilisation of $TiO_2$ is frequently carried out [21-24].

Electrophoretic deposition (EPD) is a practical method for immobilising $TiO_2$ photocatalysts as it enables rapid sample fabrication from suspensions of low solids loading [25-27]. The current work examines the effects of firing conditions on the microstructure and performance of thick films prepared by anodic electrophoretic deposition of anatase $TiO_2$ onto graphite substrates. As carbon has been reported to enhance the anatase to rutile phase transformation and lower the band gap in $TiO_2$[28-30], this method of fabrication may improve photocatalytic performance by yielding bi-phasic $TiO_2$ at lower temperatures and through carbon doping of the photocatalyst layer.

## 2. Experimental Procedure

### 2.1. EPD

Thick $TiO_2$ films were prepared through anodic EPD from acidic aqueous suspensions adjusted to pH~ 3 using oxalic acid. As reported elsewhere, the use of oxalic acid imparts a negative zeta potential to $TiO_2$ particles in suspension and thus facilitates anodic EPD from acidic aqueous suspensions with lower levels of water electrolysis [31]. Using a solids loading of 1%, a deposition voltage of 10 V and a deposition time of 10 min, $TiO_2$ anatase powder (>99%, Merck Chemicals) with a BET evaluated surface area of ~10 $m^2g^{-1}$, was deposited on 25 x 25 x 2 mm graphite substrates (GrafTech International, Ohio, USA). The average density of 8 thick films prepared was evaluated to be 64.5 $gm^{-2}$ with a standard deviation of 12.2 $gm^{-2}$ and from cross sectional examination the thickness was found to be ~80 μm.

### 2.2. Sintering

Anatase films deposited on graphite substrates were fired in air using an electric muffle furnace in the range 500-700 ºC. Graphite substrates were completely burnt off at 700ºC when fired in air, while samples fired at lower temperatures exhibited poor adhesion and substrate deterioration. Samples fired in nitrogen were fabricated in a tube furnace at temperatures 500-900 ºC with high purity nitrogen flowing through the tube at 1 l min$^{-1}$. Subsequent to firing, no substrate deterioration was observed in nitrogen fired samples and films showed good adhesion to substrates, although some loosely adhered particles were present.

### 2.3. Microstructural analysis

Scanning electron microscopy (SEM) and optical microscopy were employed to examine the microstructure of films synthesised in this work. SEM analysis was facilitated using a FEI Nova-230 SEM. Phase identification by laser Raman microspectroscopy was facilitated using a Renishaw inVia Raman microscope with laser excitation at 514 nm wavelength. Quantitative phase analysis by X-ray Diffraction was carried out using a Phillips MPD unit. Phase fractions were calculated from XRD peaks using the method of Spurr and Myers according to the following equation [32].

$$X_A = (1 + 1.26 \frac{I_R}{I_A})^{-1} \qquad (1)$$

In this equation $X_A$ is the phase fraction of anatase (assumed $X_A=1-X_R$) and $I_R$ and $I_A$ are respectively the intensity of the rutile (110) peak at 27.35° 2θ and the anatase (101) peak at 25.18° 2θ.

### 2.4. Bactericidal Activity

Bactericidal activity of the samples fabricated in this work was assessed by the inactivation of *Escherichia coli* (*E. coli*) AN180 (School of Biotechnology and Biomolecular Sciences, UNSW, Australia) in aerated water, a common approach to evaluating the bactericidal activity of $TiO_2$ photocatalysts. A diagram of the bactericidal reactor is shown in Fig. 1.

Bactericidal evaluation was carried out by adding 2 ml of overnight-incubated *E. coli* culture inTryptone Soy Broth (Oxoid, Basingstoke, UK) to 300 ml of autoclaved distilled water in which photocatalyst samples were placed. The system was irradiated by two 15W UV lamps with emission peaks at λ=350 nm. Using a Digitech QM1587 Light Meter, irradiance was evaluated to be 4.42 Wm-1 at the photocatalyst surface.

The destruction of bacteria was evaluated by determining the concentration of colony forming units (CFUs) in the treated water according to ASTM D5465. 1 ml aliquots of water were taken at fixed time intervals and serially diluted at 1:9 ratios in sterile 0.1% peptone water (Oxoid). Subsequently, 0.1 ml aliquots of the appropriate dilutions were spread-plated on Tryptone Soy Agar (Oxoid) and the plates were incubated for 24 hours at 37 °C. After incubation, colonies were enumerated and counts converted to $\log_{10}$ CFU/ml, representing the concentration of bacteria in the reactor water.

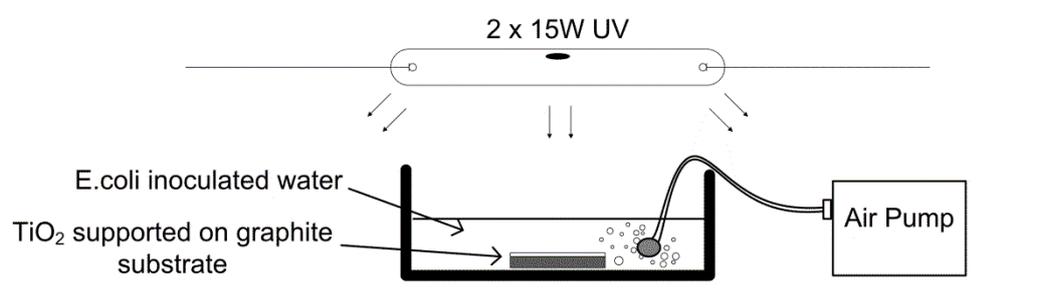

**Fig. 1. Diagram of bactericidal reactor**

### 2.5. Spectroscopy

The radiative recombination of photogenerated electron-hole pairs in the different samples was studied by examining the intensities of photoluminescence (PL) emission spectra. This was carried out by gathering diffuse spectra in the range 350-900 nm (~3.5 – 1.4 eV) using a Kimmon 20 mW 325 nm He-Cd laser in conjunction with a Renishaw inVia Raman microscope.

UV-Visible absorbance spectra were gathered to examine the shift in the absorption edge and overall absorbance between the different samples. These spectra were gathered using a Perkin-Elmer Lambda-35 UV-Vis spectrometer with a Labsphere RSA-PE-20 integrating sphere of 50 mm diameter. Scans were carried out in the wavelength range 200-700 nm with a 2 nm slit width.

## 3. Results

### 3.1. Microstructure

The prefiring microstructure of all thick films prepared in this work exhibited gas-bubble damage resulting from the parasitic process of water electrolysis as shown in Fig. 2. Holes resulting from bubble damaged ranged from ~5 to ~50 μm in size. Samples fired in air at 700ºC exhibited complete oxidation of the graphite substrate, leaving behind a fragile unsupported $TiO_2$ film. Samples fired in air at 500 and 600 ºC exhibited substrate deterioration through partial oxidation, resulting in spalling and poor adhesion of deposited films. Samples fired in nitrogen did not exhibit substrate deterioration and resulted in well adhered films showing less cracking as evident from the comparison of Fig. 3a and 3b. For samples fired at 600ºC the grain size consisted of anatase grains of ~150nm size, increasing with firing temperature, as Shown in Fig. 4. Films fired at 900 ºC showed large ~1μ coalesced grains of rutile.

Porosity was evident in films fired at all temperatures, a feature likely to be beneficial for photocatalytic applications through the increase in available surface area.

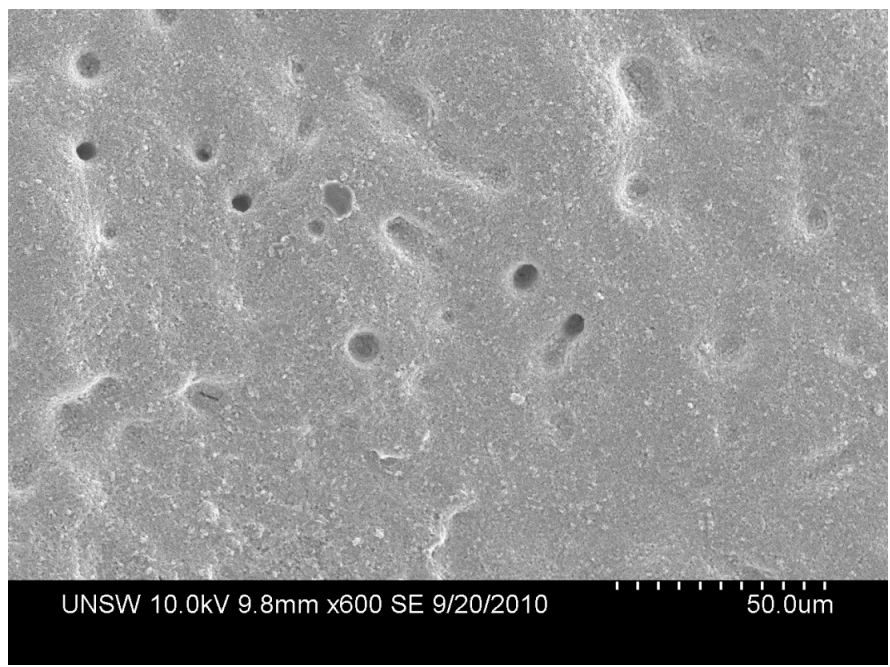

**Fig. 2. Typical microstructure showing gas-bubble damage on the surface of a thick film fired in nitrogen at 600ºC.**

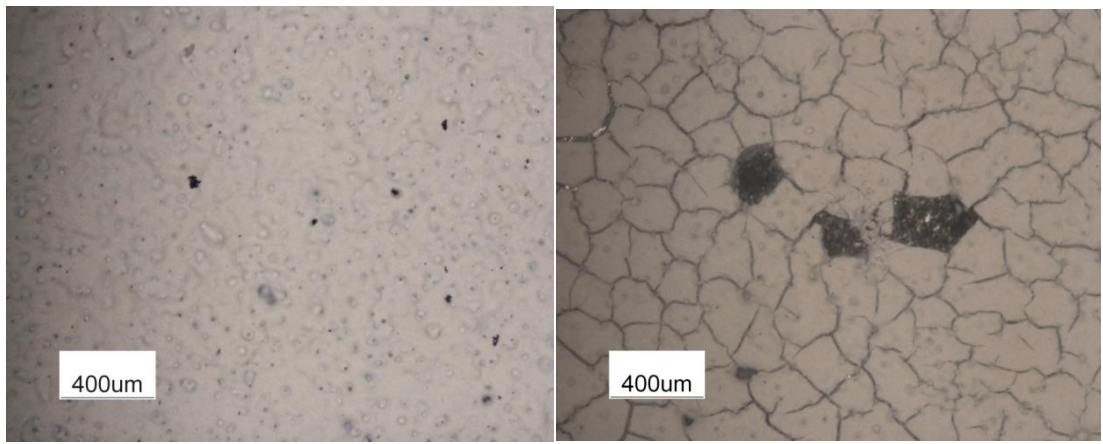

**Fig. 3.** EPD films fired at 600 ºC (a) in nitrogen (b) in air

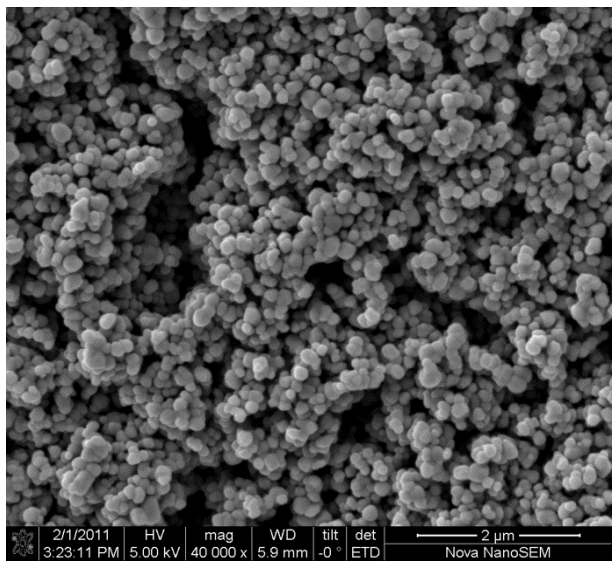

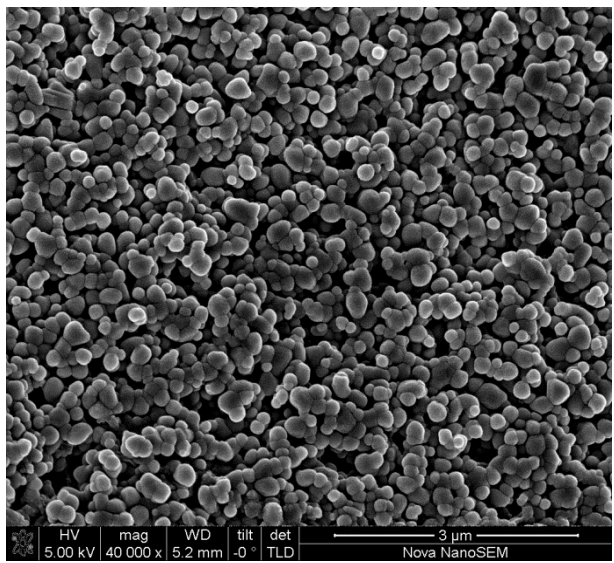

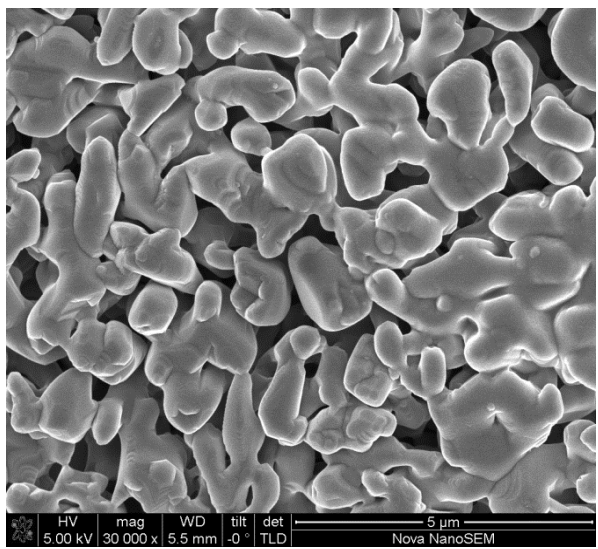

**Fig. 4. Microstructure of film fired in nitrogen at (a) 600ºC (b) 800ºC and (c) 900ºC**

### 3.2. Phase composition

XRD and Raman patterns, shown in Fig. 5. And Fig. 6. respectively, show the presence of rutile in EPD films fired in nitrogen at 800ºC with near complete transformation to rutile at 900ºC. No significant effect of firing atmosphere on phase transformation was observed as all samples fired in air showed only the anatase phase of $TiO_2$. Unsupported anatase exhibited greater thermal stability and showed only anatase peaks after firing at 800ºC in air and nitrogen.

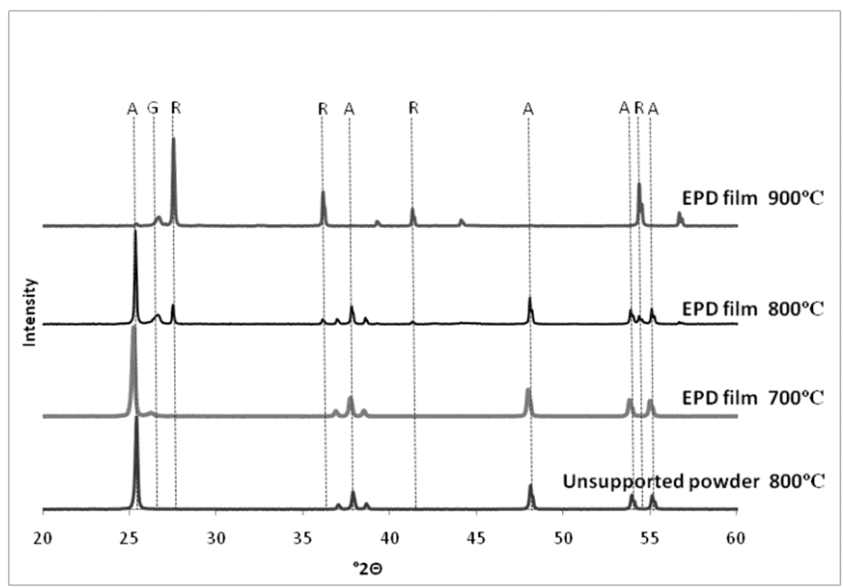

Fig. 5. XRD patterns of samples fired in nitrogen. A, R and G represent anatase, rutile and graphite respectively

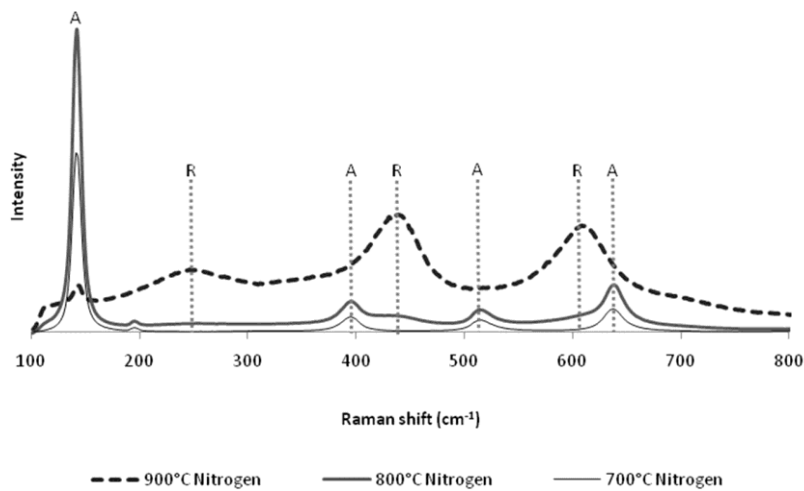

Fig. 6. Raman spectra of EPD films fired in nitrogen with anatase (A) and rutile (R) peaks marked

XRD patterns were interpreted to calculate phase fractions using the method of Spurr and Myers. The quantitative analysis of phase composition is shown in Fig. 7. The enhanced anatase to rutile transformation in graphite-supported thick films is evident from the larger rutile fraction in these samples in comparison with isothermally fired unsupported powder.

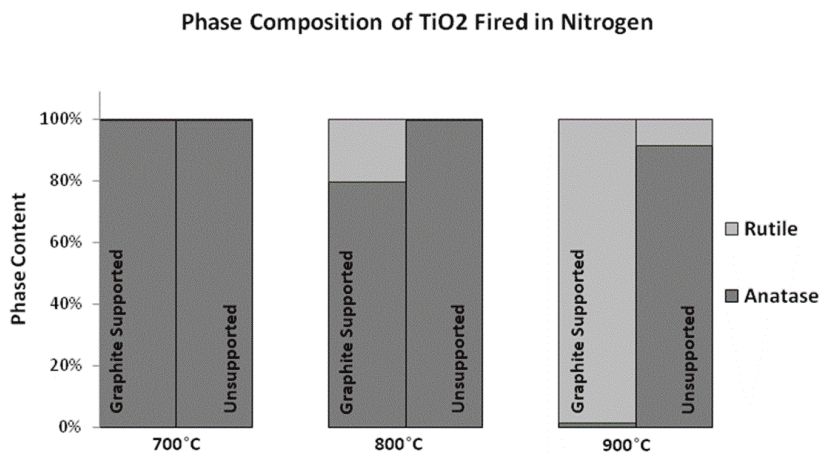

**Fig.7. Quantitative analysis of phase composition in graphite supported TiO$_2$ thick films and unsupported TiO$_2$**

### 3.3. Bactericidal Activity

The changes in the concentration of *E. coli* AN180 CFUs under UV illumination are shown in Fig. 8. An uncoated graphite substrate was used to evaluate the baseline inactivation of bacteria under UV illumination in the absence of a photocatalyst and it can be seen that only a minor decrease in CFU concentration takes place under such conditions.

Samples fired in nitrogen exhibited superior bactericidal activity than samples fired in air. Nitrogen fired TiO$_2$ thick films facilitated a > 90% inactivation rate within 20 minutes of UV irradiation while air fired samples did not achieve similar results. The effects of firing temperature on bactericidal activity are not unequivocal from the results, however it appears that sample fired at lower temperature exhibits a higher initial rate of bacteria inactivation.

Complete sterilisation of the water was not achieved within the timeframe of the experiments, rather the microbial concentration reached a sustainable level at which the rate of bacteria inactivation was offset by their natural multiplication.

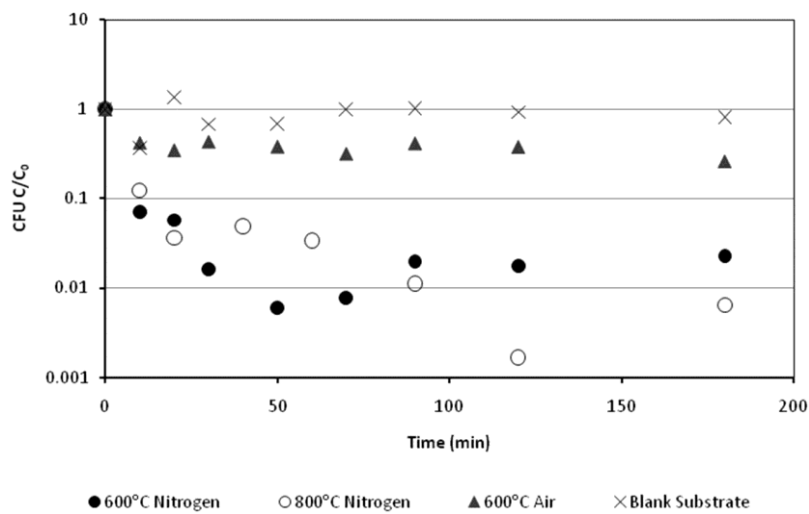

**Fig. 8.** Concentration of *E. coli* **AN180** CFUs as a function of time in bactericidal experiments using TiO$_2$ thick films

### 3.4. Spectroscopy

Photoluminescence emission spectra gathered at room temperature from different films are shown in Fig. 9. The emission peak at ~2.3 eV is consistent with the reported PL spectra of anatase [33-35]. Thick films fired in nitrogen at 600 °C exhibit higher levels of PL emission which decrease with increasing firing temperature. Consistent with reported data, the decrease in PL emission is particularly significant as the anatase to rutile transformation takes place [34, 35]. A sample fired in air exhibited lower PL emission in comparison with sample isothermally fired in nitrogen. This may be a result of lower charge carrier recombination (owing to lower excitation levels or improved electron-hole separation) or a consequence of increased scattering of the 325nm UV laser used for photoexcitation.

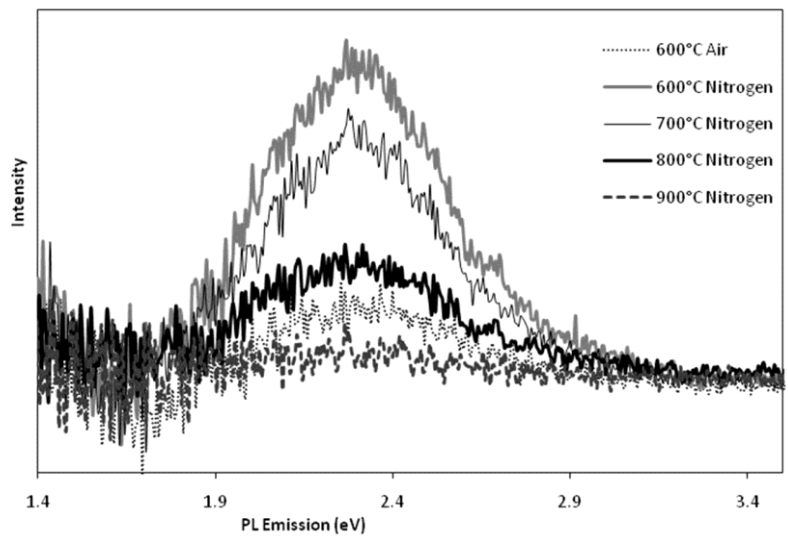

**Fig. 9. Photoluminescence emission spectra of samples excited by 325 nm irradiation**

UV-Visible absorption spectra are shown in Fig. 10. An absorption edge at around 380-390 nm corresponds to the band gap of anatase $TiO_2$ of ~3.2 eV. It can be observed that samples fired in nitrogen exhibit higher overall absorption and a more moderate slope at the absorption edge. These results cannot be interpreted to determine the photocatalytic performance of the material as increased absorption does not necessarily imply increased photogeneration of electron-hole pairs. Furthermore the differing levels of exposure of the graphite substrates bring about a shift in the absorption levels of the films. The step at 326 nm is a result of the irradiation lamp changeover at this wavelength.

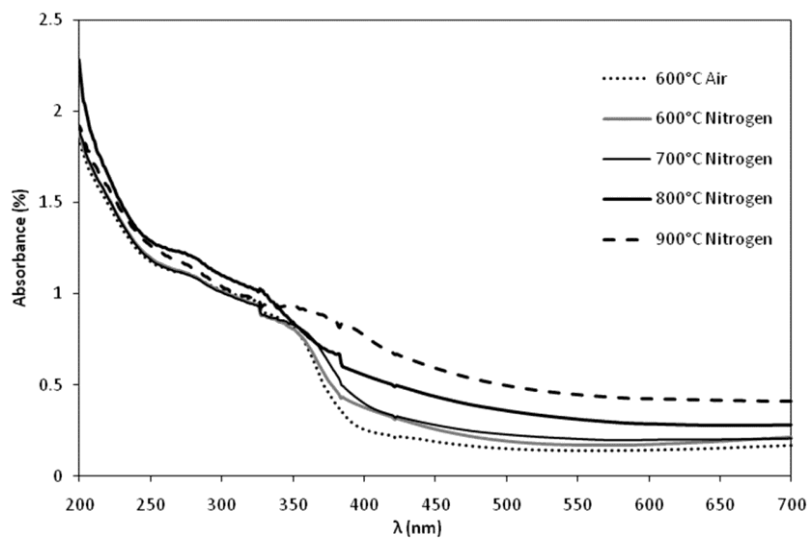

Fig. 10. UV-Visible absorption of $TiO_2$ thick films fired at different temperatures in air and nitrogen

## 4. Discussion

### 4.1. Effect of firing conditions on microstructure

Graphite is generally reported to exhibit rapid oxidation in air around 700 °C [36, 37] and thus the oxidation of the substrates fired at 700°C in air in this work was anticipated. As would be expected, the oxidation of the graphite substrate for films fired in air has a detrimental effect on the structure and adhesion of the deposited thick film. This is evident from increased spalling and cracking, shown in Fig. 3, and the low resilience of the air-fired films to abrasion, suggesting firing in air is an unsuitable treatment for EPD films on graphite substrates, even at temperatures below the ignition temperature of graphite. In contrast, EPD thick films fired in nitrogen did not show oxidation damage and exhibited superior adhesion. As shown in Fig. 4, grain size increased with increasing firing temperature, with a significant growth occurring between 800-900°C, as the phase transformation to rutile reached near completion. Significant grain growth is likely to be detrimental to the photocatalytic activity of the material due to a decrease in available surface area; however the partial transformation to rutile may be beneficial for the photocatalytic activity through improved charge carrier separation as reported elsewhere.

### 4.2. Phase composition

The anatase phase of the powder used in this work shows greater thermal stability to what is frequently reported in the literature. While anatase is typically reported to transform to rutile at temperatures between 600 and 700 °C [30, 38-41], unsupported powder in this work remained entirely in the anatase phase after firing at 800°C. Similar commercially available anatase has shown thermally stable anatase phase in other work [42, 43], this thermal stability is likely to be due to low levels of silica impurities in the raw material [30]. The presence of the graphite substrate promotes the anatase to rutile transformation. This promotion of the phase transformation, illustrated in Fig. 5. And Fig. 7., is most likely due to the increase in oxygen vacancies in the anatase lattice as reported elsewhere [30]. In a non-oxidising atmosphere, the carbon in the graphite substrate may cause a partial reduction of the $TiO_2$ film giving rise to the formation of oxygen vacancies and $Ti^{+3}$ species, the presence of which enhances the anatase to rutile phase transformation by easing the atomic rearrangement involved in this transformation [44].

### 4.3. Bactericidal activity

From Fig. 8. it can be seen that samples fired in nitrogen exhibited superior photocatalytic performance in comparison with air fired material. Higher photocatalytic activity of $TiO_2$ fired in nitrogen has been reported previously [45]. A likely explanation of the enhanced photocatalytic performance observed in samples fired in nitrogen is that this treatment enables the diffusion of nitrogen atoms from the firing atmosphere and carbon atoms from the substrate, into the $TiO_2$ lattice which facilitate an increase in exciton photo-generation. As reported elsewhere [28, 29, 46], the substitution of oxygen with nitrogen and carbon atoms in $TiO_2$ gives rise to new valence states and thus increases the optical response by a decrease in the band-gap of the material. It has also been reported that the substitution of carbon and nitrogen in place of oxygen in $TiO_2$ reduces charge carrier recombination [28]. Conversely, in samples fired in air, bactericidal activity was low, showing only moderate activity relative to an uncoated graphite substrate. The lower photocatalytic activity of air fired samples may be a result of substrate oxidation which inhibited the diffusion of carbon into the $TiO_2$ lattice and brought about deterioration in the quality of EPD films which resulted in a loss of photocatalyst in the bactericidal reactor.

The effect of increasing firing temperature on the bactericidal activity is not unequivocally clear from the results shown in Fig. 8. It appears the material fired at a lower temperature brings about a more rapid initial bacterial inactivation, however the sample fired at 800°C exhibits a lower final CFU concentration. The differences in bactericidal performance between the two samples fired in nitrogen at different temperatures are not of a significant magnitude, and the ambiguity may result from the mixed effect of grain size and phase composition. The lower-temperature fired material exhibits higher surface area owing to the finer grain size visible in Fig. 4, however the material fired at 800°C shows a secondary rutile phase, potentially improving charge carrier separation and consequently improving photocatalytic activity [13-17]. Furthermore the material fired at 800°C may exhibit greater levels of carbon and nitrogen diffusion in the $TiO_2$ lattice, giving rise to a lower band-gap.

In general, the bactericidal activity observed in this work was notably low in comparison with results reported elsewhere [19, 20, 47], and no complete sterilisation was achieved. The comparatively low rates of *E. coli* inactivation evident from Fig. 8. are likely to be the result of small catalyst area in comparison with the reactor dimensions, low irradiance levels, the use of air sparging rather than pure oxygen sparging, and the surface area of the commercially available material used which is lower than catalysts used in other work. Irradiance levels in the reactor used in this work were measured at 4.42 $Wm^{-1}$ while the UV irradiance of sunlight is up to 50 $Wm^{-1}$ [48]. This suggests that greater efficiencies can be achieved using natural solar irradiation rather than illumination by a UV lamp. The deposition of higher surface area $TiO_2$ powder in conjunction with nitrogen firing may yield improved performance than the samples prepared in this work.

### 4.4. Spectroscopy

The photoluminescence spectra of thick films fired in nitrogen shown in Fig.9 exhibit a decrease in PL emission intensity with increasing firing temperature. Increased PL emission in $TiO_2$ results generally from increased radiative recombination of excitons [49, 50] and may indicate enhanced photo-generation of these electron-hole pairs, a faster rate of their recombination or a combination of both of these phenomena [51-53]. Intensity of PL emission may also vary as a result of surface properties and resultant variation in the scattering of the photoexciting UV [53]. Consequently, similar to UV-Vis absorbance, PL emission intensity cannot be used to directly infer photocatalytic activity.

The spectra in Fig. 9 show higher levels of PL emission from samples fired in $N_2$ in comparison with a sample fired in air. An increase in PL emission in $TiO_2$ fired in an oxygen deficient atmosphere has been previously reported as a result of increased oxygen vacancies[34]. Additionally, the PL spectra from the sample fired in air at 600 °C may be diffuse owing to increased surface roughness resulting from oxidation of the substrate and consequent deterioration of the thick $TiO_2$ film.

The similar levels photocatalyzed inactivation of *E. coli* exhibited by samples fired at 600 °C and 800 °C in nitrogen suggest that the lower PL emission intensity of the sample fired at 800 °C in nitrogen can be attributed ,at least partly, to improved charge carrier separation in this sample resulting from a mixed anatase-rutile phase composition. If the lower PL emission intensity in the sample fired at 800 °C was purely a result of lower levels of excitation, this material would exhibit markedly poorer photocatalytic activity in the inactivation of bacteria. Conversely if the lower PL emission intensity was solely the result of improved charge carrier separation, this material would be expected to exhibit noticeably higher activity.

UV-Visible spectra shown in Fig. 10 show a more moderate slope at the adsorption edge in nitrogen doped samples with higher overall absorption. These spectra are consistent with the aforementioned formation of new valence states by nitrogen and/or carbon diffusion and the resultant increased optical response [54]. A further increase in overall absorption and broadening of the UV-Vis absorption spectra can be seen as a result of rutile formation. This is consistent with reports that the formation of rutile at low levels is sufficient to shift the absorption edge of $TiO_2$ to higher wavelengths[24].

## 5. Conclusions

Porous thick films of $TiO_2$ can be fabricated on graphite substrates by using a method of anodic aqueous EPD. When such fabrication methods are combined with firing in a nitrogen atmosphere a well adhered film exhibiting enhanced photocatalytic activity can be obtained.

The anatase to rutile transformation is enhanced in thick films on graphite substrates as a result of increased levels of oxygen vacancies created by the diffusion of carbon atoms into the $TiO_2$ lattice. The diffusion of carbon and nitrogen into the $TiO_2$ lattice may also explain the improved photocatalytic activity of material fired in nitrogen in comparison with air fired material. A mixed phase composition, achieved by firing at 800°C in nitrogen, further enhances photocatalytic activity through improved charge carrier separation.

## Acknowledgements

The authors acknowledge access to the UNSW node of the Australian Microscopy and Microanalysis Research Facility (AMMRF) and the assistance of Anne Rich of the spectroscopy lab at the Mark Wainwright analytical centre at UNSW.